\documentclass{iau}
\usepackage{graphicx}
\usepackage{caption}
\usepackage{hyperref}

\title[Asteroids in LSST] 
{Asteroid Discovery and Characterization with the Large Synoptic Survey Telescope}

\author[R. L.  Jones, M. Juri\'{c}, \v{Z}. Ivezi\'{c}]   
{R. Lynne Jones$^1$, Mario Juri\'{c}$^2$, 
 \and \v{Z}eljko Ivezi\'{c} $^3$}

\affiliation{$^1$University of Washington, email: {\tt ljones@astro.washington.edu} \\
$^2$University of Washington, email: {\tt mjuric@astro.astro.washington.edu} \\
$^3$University of Washington, email: {\tt ivezic@astro.astro.washington.edu}}

\pubyear{2015}
\volume{318}  
\jname{Asteroids: New Observations, New Models}
\editors{S. Chesley, A. Morbidelli, R. Jedicke \&
  D. Farnocchia eds.}
\begin{document}

\maketitle

\begin{abstract}
The Large Synoptic Survey Telescope (LSST) will be a ground-based,
optical, all-sky, rapid cadence survey project with tremendous
potential for discovering and characterizing asteroids.

With LSST's large 6.5m diameter primary mirror, a wide 9.6 square
degree field of view 3.2 Gigapixel camera, and rapid observational
cadence, LSST will discover more than 5 million asteroids over its ten
year survey lifetime. With a single visit limiting magnitude of 24.5
in $r$ band, LSST will be able to detect asteroids in the Main Belt
down to sub-kilometer sizes.  The current strawman for the LSST survey
strategy is to obtain two visits (each `visit' being a pair of
back-to-back 15s exposures) per field, separated by about 30 minutes,
covering the entire visible sky every 3-4 days throughout the
observing season, for ten years.

The catalogs generated by LSST will increase the known number of small
bodies in the Solar System by a factor of 10-100 times, among all
populations. The median number of observations for Main Belt asteroids
will be on the order of 200-300, with Near Earth Objects receiving a
median of 90 observations. These observations will be spread among
$ugrizy$ bandpasses, providing photometric colors and allow
sparse lightcurve inversion to determine rotation periods, spin axes, and shape information.

These catalogs will be created using automated detection software, the
LSST Moving Object Processing System (MOPS), that will take advantage
of the carefully characterized LSST optical system, cosmetically
clean camera, and recent improvements in difference imaging. Tests
with the prototype MOPS software indicate that linking detections (and thus
`discovery') will be possible at LSST depths with our working
model for the survey strategy, but evaluation of MOPS and improvements
in the survey strategy will continue. All data products and software created by
LSST will be publicly available.
\end{abstract}

\firstsection 

\section{Introduction}

The Large Synoptic Survey Telescope (LSST) is a next-generation survey
project, coupling a world-class telescope facility with cutting-edge
data management software and calibration efforts. Its primary science
drivers are to constrain dark matter and dark energy, to map the Milky
Way and Local Volume, to catalog the Solar System, and to explore
the transient optical sky. The catalogs generated by LSST during its
ten years of operation will enable a multitude of science
investigations beyond these primary science drivers, many of which are
explored in the LSST Science Book (\cite{scibook}). 

The inventory of the Solar System is one of the primary science
drivers for LSST. Fulfilling this science goal will involve
discovering millions of minor planets, increasing the number of known
objects in every small body population by a factor of 10 to 100 above
current levels. Many of these objects will receive large numbers
($>100$) of observations, over a long time span (several years) and with
extremely accurate astrometry (10mas errors), resulting in highly
accurate orbits suitable for a wide range of theoretical studies or
for targeted follow up observations for specific purposes (such as
spectroscopy or occultation studies.

These large number of observations will also provide the basis for
sparse lightcurve inversion, which requires at least 100 observations
over a range of phase angles. It will be possible to determine
the spin states and shapes for thousands of Main belt
asteroids. Frequent observations, spread among a wide range of times
and at variety of different points along each object's orbit, are also
ideal for detecting activity, either collisionally-induced activity or
surface activity induced by volatile outgassing.

Each object will obtain observations in different filters, primarily $g$,
$r$, $i$ and $z$ but also $u$ and $y$, with photometric calibration of
each measurement accurate to 10mmags (\cite{lsstoverview}). This will enable study of the
composition of these objects. Adding color information also
provides statistical constraints on the albedos of the objects,
allowing a tighter estimate of the diameters and thus size
distribution of the population. With combined color and orbital
information, identification of collisional families becomes more
robust. (See the Solar System chapter from \cite{scibook} for further discussion of these
topics). Table \ref{table1} provides a summary of the expected number
of objects in each population, as well as their typical arc length and
number of observations.

\begin{table}
\begin{center}
\caption{Summary of small body populations observed with LSST}
\label{table1}
 {\scriptsize
  \begin{tabular}{|l|c|c|c|c|}\hline
Population & Currently known$^1$ &  LSST discoveries$^2$ &
Num. of obervations$^3$
    & Arc length (years)$^3$ \\ \hline
Near Earth Objects \\(NEOs) & 12,832 & 100,000 & (H$\leq$20) 90 & 7.0 \\
    \hline
Main Belt Asteroids \\(MBAs) & 636,499 & 5,500,000 & (H$\leq$19) 200 &
                                                                     8.5
    \\ \hline
Jupiter Trojans & 6,387 & 280,000 & (H$\leq$16) 300 & 8.7 \\ \hline
TransNeptunian and \\ Scattered Disk Objects \\ (TNOs and SDOs) & 1,921 &
                                                                    40,000
                                                                          &
                                                                            (H$\leq$6)
                                                                            450
    & 8.5 \\ \hline
 \end{tabular}
  }
 \end{center}
\vspace{1mm}
 \scriptsize{
 {\it Notes:}\\
  $^1$As reported by the MPC (May 2015).
  $^2$Expected at the end of LSST's ten years of operations.
  $^3$Median number of observations and observational arc length for the brightest objects near
  100\% completeness (as indicated). }
\end{table}

Construction for LSST is ongoing, with first light scheduled for 2020,
a scientific commissioning program following in the next year, and the
start of survey operations in 2022. Details of LSST operations are
currently being examined. In particular, the survey strategy continue
to be analyzed up to and during operations in order to maximize the
science return across the wide variety of goals for LSST. In this
proceedings, we will describe the planned LSST configuration, and
expected LSST performance in discovering and characterizing Near Earth
Objects (NEOs) and Main Belt asteroids (MBAs), then present software
tools that can aid the planetary astronomy community in extending
this analysis.

\section{The LSST telescope}

The primary science goals for LSST drive the design of the telescope
and camera. The choice of telescope mirror size, field of view,
filters and typical exposure times combine to achieve the desired
single image depth, coadded image depth, number of repeat visits, 
visit distribution among filters, and survey footprint.

The resulting final design is an optical telescope with
$ugrizy$ filters and a primary mirror of 8.4m in
diameter (the effective diameter is 6.5m after accounting for obscuration
and vignetting). The telescope has a fast f/1.2 focal ratio; together
with the 3.2 Gigapixel camera, this provides a 9.6 square degree field
of view with 0.2 ''/pixel platescale. Short exposures and a rapid
survey strategy covering the entire visible sky every three to four
nights in multiple filters complete the basic strategy to meet these
science goals.

The details of the observing strategy will be discussed further in
Section~\ref{surveystrategy}, but at the base of the cadence is the
pair of back-to-back 15 second exposures that make up a 30 second
`visit'. For most purposes, this 30 second visit can be considered the
equivalent exposure time for LSST; the back-to-back `snaps' will be
processed separately to help reject cosmic rays (and could be used to
help determine velocity direction for trailing moving objects), but
the images will be combined for most purposes and individual image
depths correspond to the 30 second visit $5\sigma$ point-source
limiting magnitude. This drives further design choices for the
telescope; in order to maintain a high duty cycle, the camera readout time
is only 2 seconds per exposure and the slew-settle time between
nearby fields is only 5 seconds per visit.

\begin{figure}[tb]
\begin{minipage}{.5\textwidth}
\begin{center}
\includegraphics[width=0.9\linewidth]{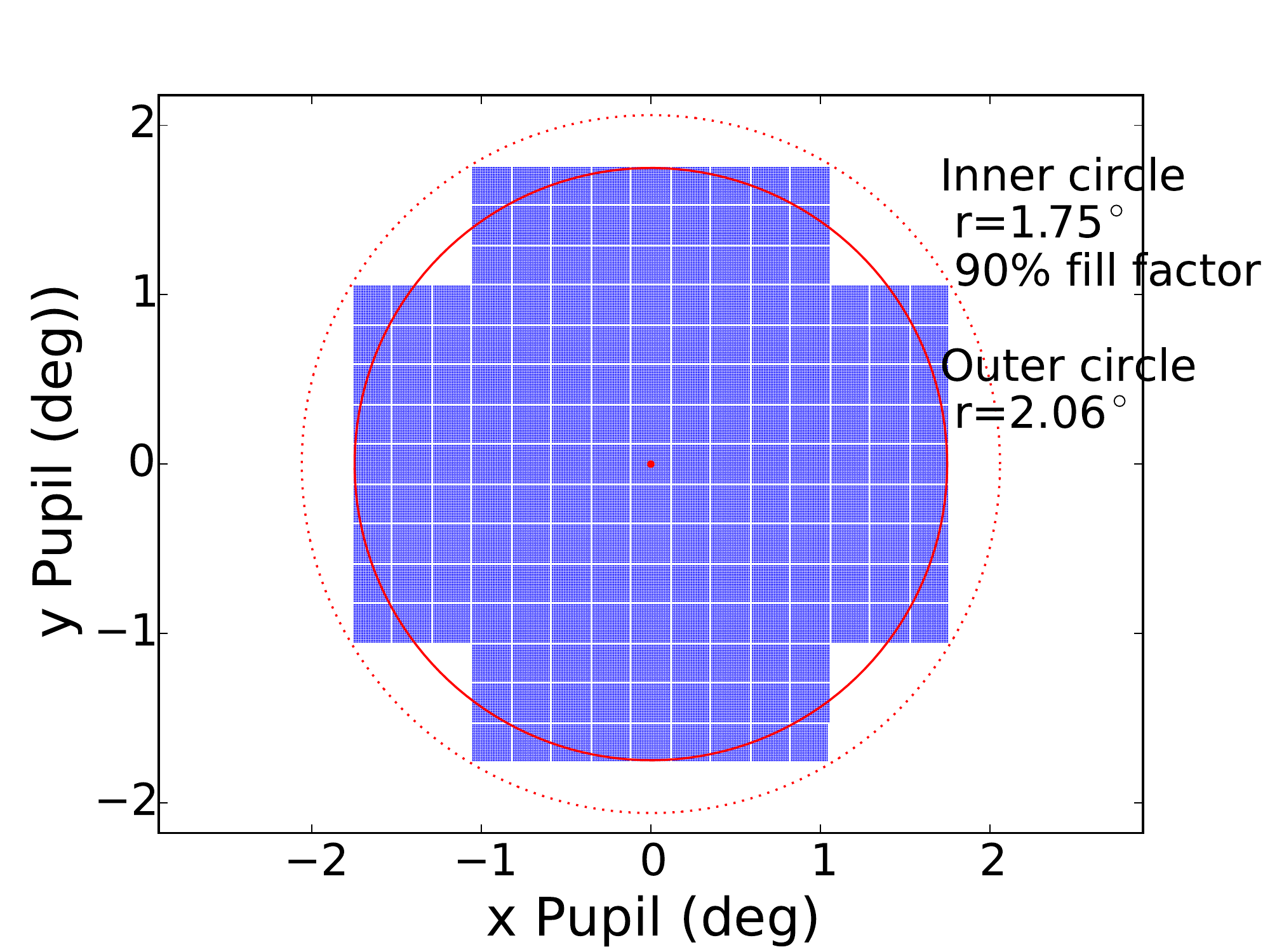}
\captionsetup{width=0.9\linewidth}
\caption{Layout of the LSST focal plane. The solid circle indicates
  the inscribed circular field of view (3.5$^\circ$ diameter). The
  plotted points indicate active silicon.\label{focalplane}}
\end{center}
\end{minipage}
\begin{minipage}{.5\textwidth}
\begin{center}
\includegraphics[width=0.9\linewidth]{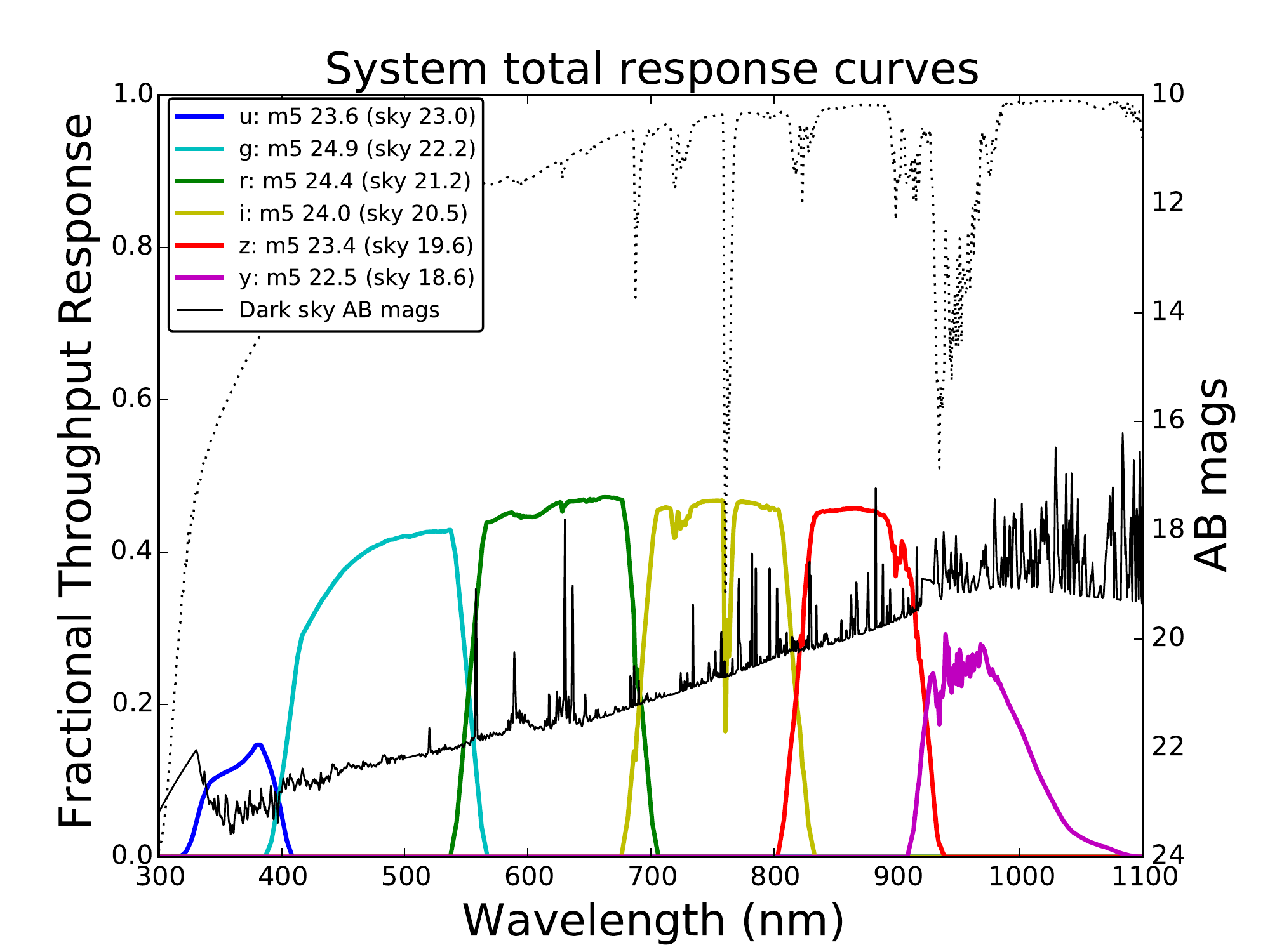}
\captionsetup{width=0.9\linewidth}
\caption{Expected LSST throughput response in $ugrizy$, including
  an atmospheric throughput curve (the dotted line). The
  expected dark sky brightness in AB magnitudes is also shown (the
  thin black line).
\label{throughputs}}
\end{center}
\end{minipage}
\end{figure}

The fill factor of the camera is 90\%, counting active silicon within
a $3.5^\circ$ diameter circle inscribed in the field of view; the fill
factor counting only chip gaps but over the entire (non-circular)
focal plane is slightly higher, but similar. See
Figure~\ref{focalplane} for an illustration of the focal plane.

On-site monitoring has provided information on the expected free
atmosphere FWHM and sky brightness (see Table~\ref{table2}). The
telescope hardware is expected to contribute an additional 0.4'' to
the delivered seeing. The expected dark sky skybrightness is generated using
detailed sky spectra obtained elsewhere (\cite{patat}), modified to
match broad-band sky brightness measurements reported from Cerro
Pachon and other nearby sites.

Expected throughput curves for each component of the hardware system
are maintained by system engineering (see the github
repo\footnote{\url{http://github.com/lsst-pst/syseng_throughputs}} for
latest values).  These are based on data from prototype sensors and
the expected performance of the mirrors and filters and lenses,
including broadband coatings and loss estimates due to condensation
and contamination. The throughput curves for each filter are
illustrated in Figure~\ref{throughputs}.

Combining all of the information above, we can calculate the
expected five-sigma point-source limiting magnitudes for LSST, under
fiducial seeing and dark sky conditions - see Table~\ref{table2}.

As LSST continues to move toward operations, the expected values for each
of these components will be replaced by `as delivered'
versions. Up-to-date values will be maintained in the github
repositories and reported in the LSST Overview Paper (\cite{lsstoverview}).

\begin{table}[tbh]
\begin{center}
\caption{$ugrizy$ $5\sigma$ point source limiting magnitudes$^1$}
\label{table2}
 {\scriptsize
  \begin{tabular}{|l|c|c|c|c|c|c|}\hline
 & $u$ & $g$ & $r$ & $i$ & $z$ & $y$ \\ \hline
Median atmospheric IQ$^2$  & 0.66 & 0.61 & 0.56 & 0.53 & 0.50 & 0.48 \\
    \hline
Dark sky brightness (mag/sq'')$^3$ & 23.0 & 22.2 &  21.2 & 20.5 & 19.6 & 18.6
    \\ \hline
$5\sigma$ limiting magnitude & 23.6 & 24.9 & 24.4 & 24.0 & 23.4 & 22.5
    \\ \hline
 \end{tabular}
  }
 \end{center}
\vspace{1mm}
 \scriptsize{
 {\it Notes:}\\
  $^1$Please see the LSST Overview Paper (\cite{lsstoverview}) for
  updated values.
  $^2$Based on Cerro Pachon site monitoring.
  $^3$Based on dark sky spectra convolved with the LSST bandpasses,
  validated with site monitoring data from Cerro Pachon.}
\end{table}

\section{LSST data management}

LSST will acquire millions of images -- on the order of 2.5 million
visits, each consisting of a pair of exposures. The LSST Data
Management (DM) software pipeline has the task of turning these images
into catalogs enabling the primary science goals. In general these
catalogs can be thought of as falling into three categories: Level 1,
Level 2 and Level 3.

Level 1 data products are created during nightly processing. The
images in each visit are combined to reject cosmic rays, then
subtracted from a template image created from previously acquired
imaging (typically 6 months to a year earlier). The detections
measured in each difference image correspond to transients, variables,
moving objects, and artifacts. These outputs will be run through
machine learning algorithms to help reject artifacts. The resulting
detections, along with relevant information from existing catalogs
such as identification of known variable stars or the location of
nearby background galaxies, will be released within 60 seconds of the
end of each visit as the LSST Alert stream.

In addition, these difference image catalogs (after removing known
variable stars) will be used to feed the LSST moving object processing
system (MOPS). MOPS will link detections from different visits within
a night into tracklets, combine these with tracklets from
other nights into tracks, and finally fit the tracks with orbits; it will also extend
known orbits with new detections of these objects. These moving object
catalogs will updated and released on a daily basis.

The Alert stream and the moving object catalogs (the linked orbits and
their individual detections) make up the Level 1 data products. It is
worth noting that moving objects which are measurably trailed in any
individual visit will be clearly identifiable in the Alert stream as
such; very fast-moving objects thus have an additional discovery
avenue via Alerts.

Level 2 data products are created during yearly data processing and
include a more precise level of calibration in photometry and
astrometry. During the yearly data processing, all existing images
will be reprocessed using the most recent software release (including
reprocessing these images through MOPS, likely using slightly
improved templates for image differencing). These
data release catalogs will reach 10mmag absolute photometric accuracy
and 10mas absolute astrometric accuracy. The increased accuracy is
possible due to various algorithms that compute global solutions;
these are not run during nightly data processing.

Level 3 data products indicate data products resulting from
independently written (non-project) software, created using LSST data
access center compute resources. These data products will typically be
generated using extensions to the LSST DM software, and may or may be
publicly available depending on the user. Publicly available Level 3
data products which prove particularly useful could become fully
federated with LSST databases.

The LSST DM pipeline will be entirely open source and publicly
available. The various repositories that make up the DM software stack
can currently be found on github\footnote{
\url{http://github.com/lsst}}; more information about the stack and
instructions for installing the LSST software stack can be found at
\url{http://dm.lsst.org}. Details of the data products (images and
catalogs) are defined in the LSST Data Products Definitions Document
(DPDD)\footnote{
\url{http://www.lsst.org/content/data-products-definition-document}}
. All LSST data products will be immediately publicly available to
institutions with data rights.

\section{LSST survey strategy}
\label{surveystrategy}

The basic parameters of LSST -- telescope
size, field of view -- have been fixed. In
addition, given the survey length, visit exposure time, and constraints on the survey
footprint, an approximate outer envelope of the survey characteristics can be
estimated: the survey has about 2.5 million visits to
distribute over about 25,000 square degrees for all survey fields,
with about 825 visits per field in the main survey footprint
($\approx18,000$ sq deg) to distribute among $ugrizy$ filters. Most
fields in the main survey footprint can be observed twice per night
(with an interval of about 30 minutes) every three to four days, on an
ongoing basis over their observing season, repeating for ten
years. This is the strawman LSST observing strategy at present.

However, the details of the observing cadence have {\bf not} yet been
fixed. For example, instead of distributing visits fairly evenly in time for all
fields over all ten years, a variant may be to
concentrate a subset of those visits for some fields into a shorter
period of time (a `rolling cadence'). One option that may be interesting for
studying solar system objects could include taking more frequent
observations for fields near opposition and then reducing the number
of observations for fields away from opposition. The process of
optimizing the survey strategy in terms of cadence is just starting to
get underway.

\begin{figure}[tb]
\centering
\includegraphics[width=0.8\textwidth]{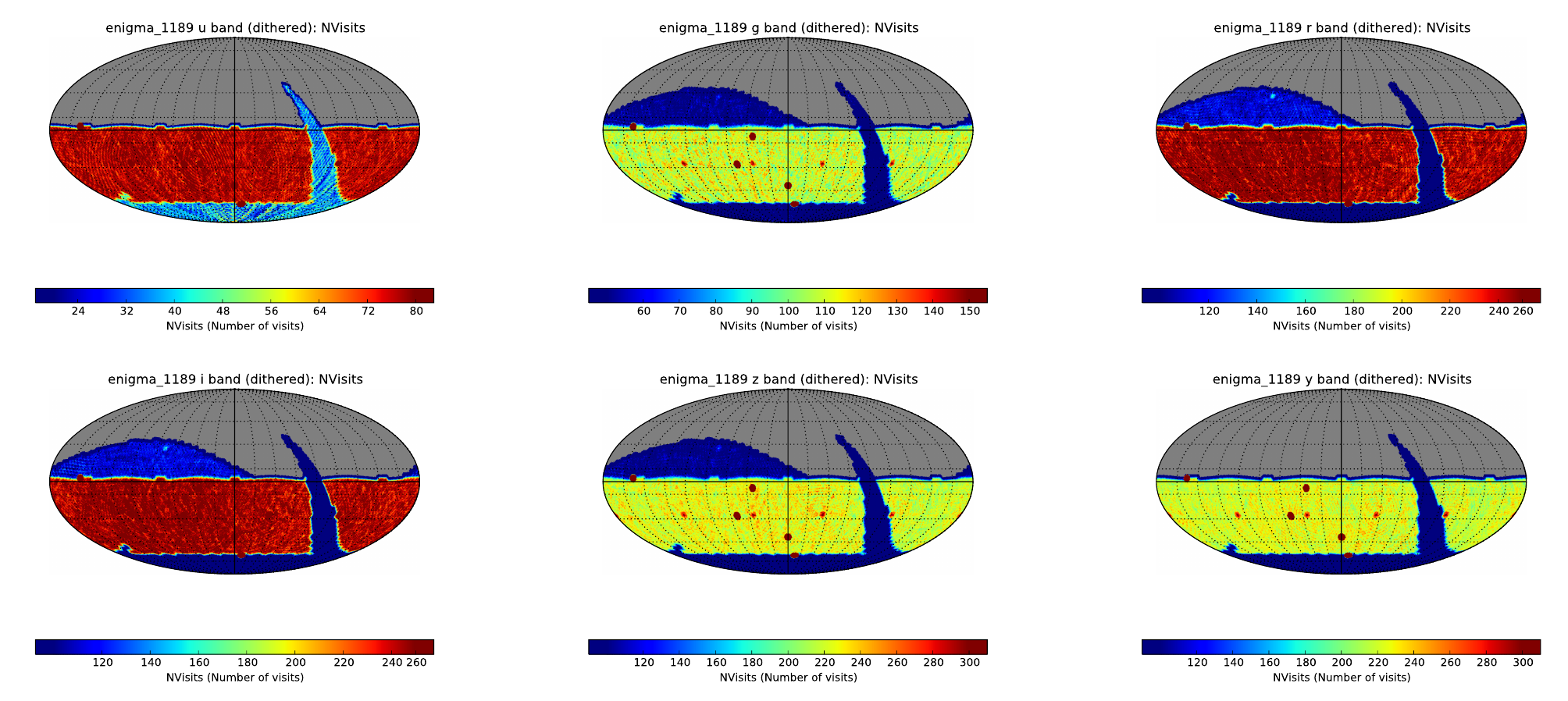}
\caption{The distribution of visits across the LSST survey footprint,
  in a sample OpSim simulated survey. The main survey covers the area
  from $-65^\circ<$ Dec $<5^\circ$, excluding a small area around the
  galactic plane. The area from Dec=$5^\circ$  up to $10^\circ$ north
  of the ecliptic is covered in an additional observing program; other
  programs cover the South Celestial Pole and the Galactic Plane.
\label{footprint}}
\end{figure}

LSST has several tools to help this process of survey strategy
optimization. The first is the LSST Operations Simulator (OpSim)
(\cite{opsim}), which combines a realistic weather history and a
high-fidelity telescope model with a scheduler driven by a set of
proposals that attempt to parametrize a basic observing strategy ({\it
e.g.}, a proposal for the main survey footprint that specifies the
main survey footprint, skybrightness and seeing limits, number of
visits desired in each filter, and the time window between pairs of
visits in each night). The output of OpSim is a simulated pointing
history, complete with observing conditions and individual visit
limiting magnitudes, that demonstrate how LSST might survey the sky. A
simple visualization of an OpSim run is shown in
Figure~\ref{footprint}; this also shows the footprint for the
survey in various proposals and filters.

\begin{figure}[tb]
\centering
\includegraphics[width=0.4\textwidth]{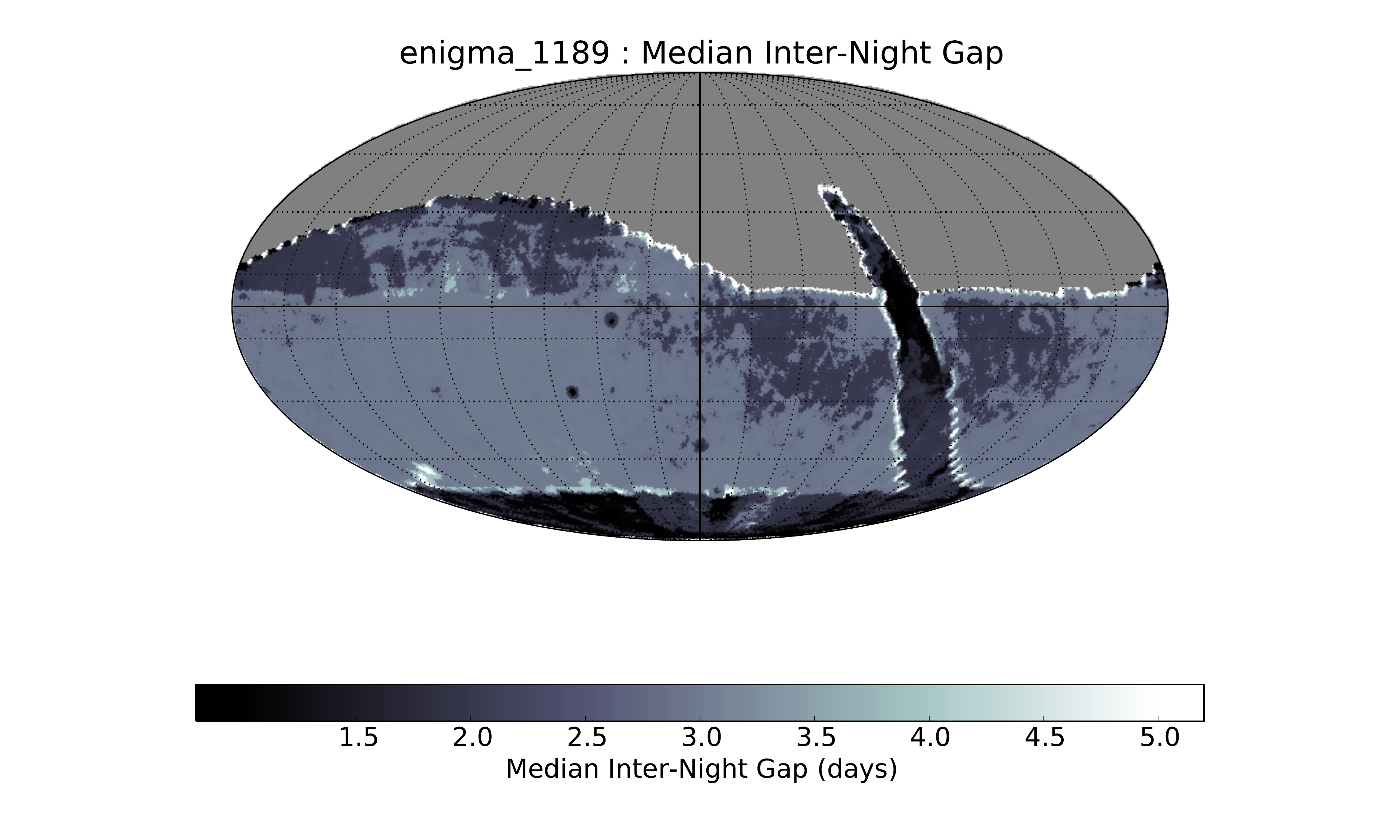}
\includegraphics[width=0.4\textwidth]{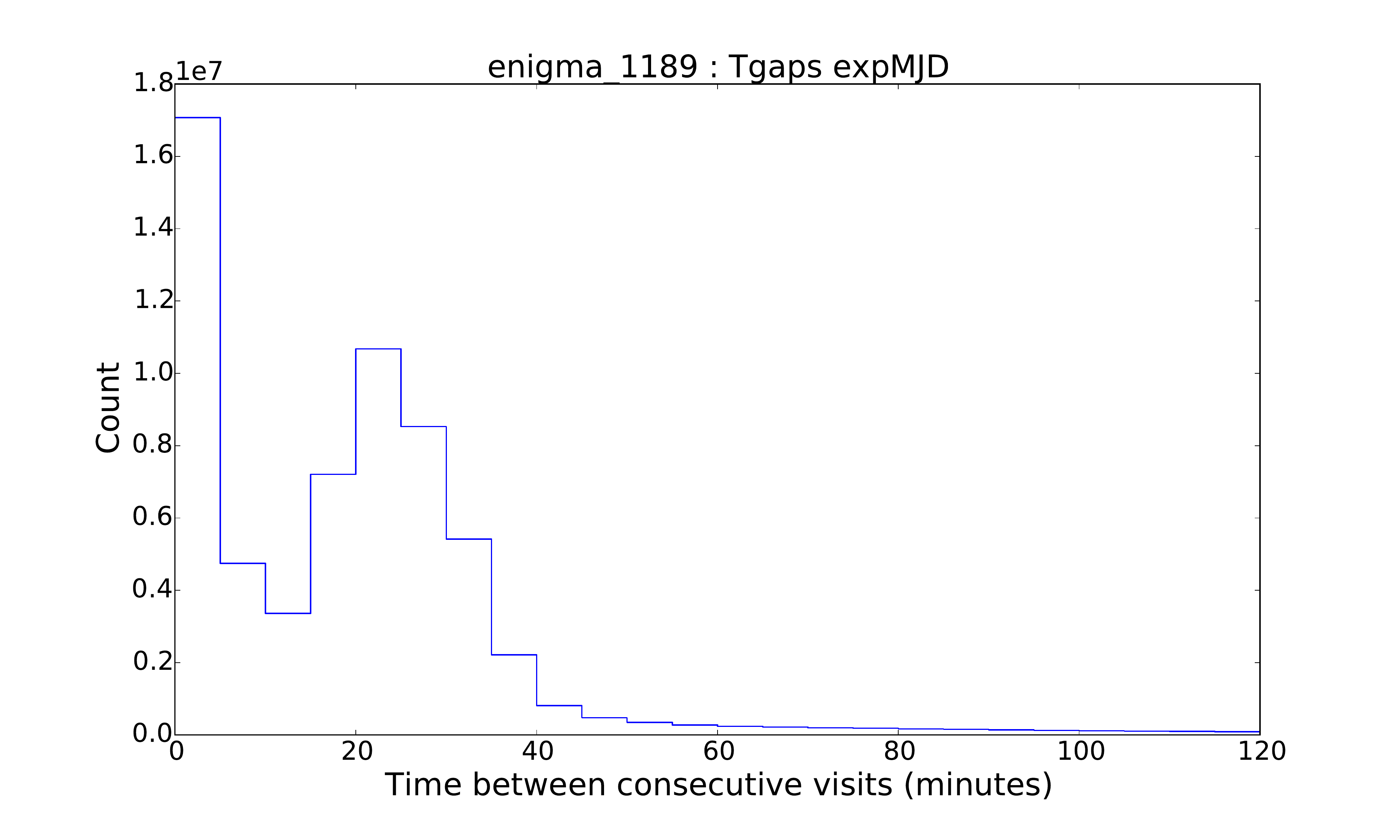}
\caption{Left: the median number of nights between consecutive
  visits to a field, for an OpSim simulated survey. Right: A histogram
  of revisit times, within each night.
\label{time}}
\end{figure}

The second tool is a user-extensible python package called the LSST
Metrics Analysis Framework (MAF) (\cite{maf}). MAF was created to help
analyze OpSim outputs. Using MAF, it is simple to write short pieces
of python code (`metrics') that can be plugged into the framework to
evaluate some aspect of OpSim. By collecting these metrics from a wide
representation of the astronomical community, we can evaluate OpSim
surveys created with a variety of scheduler configurations and
maximize science return from LSST across a wide range of science
goals. An example of using MAF to evaluate the median time between
revisits at each point on the sky is given in Figure~\ref{time}.

OpSim and MAF are open-source software packages,
provided as part of the LSST Systems Engineering Simulations
effort. Instructions for installing them are available at
online\footnote{
\url{https://confluence.lsstcorp.org/display/SIM/Catalogs+and+MAF}}. 

\section{Evaluating the LSST survey strategy for Solar System science}

The LSST Metrics Analysis Framework (MAF) can also be used to evaluate the
performance of OpSim simulated surveys with respect to Solar System
science goals. MAF will allow a user to specify a population of moving
objects (by providing their orbital parameters), specify a particular
OpSim survey, and then generate their simulated observations. MAF uses
the open-source package OpenOrb (\cite{oorb}) in generating the
ephemeris information for these simulated observations.

In many cases, the input population of moving objects can be
small (on the order of a few thousand); MAF is able to clone the resulting
detections over a range of $H$ magnitudes, so that 
metrics can be evaluated over a wide range of $H$ while only
using a relatively small set of orbits. MAF tests using 10,000 MBAs
produced identical metric results as tests using 2,000 MBAs;
statistically, this method of cloning the input population is 
adequate for most purposes.

As MAF generates the detection lists for each object, the reference
$H$ value in the orbit file is used to generate an apparent $V$ band
magnitude, then the (optionally user-assigned) spectrum is used to
generate a magnitude in the LSST bandpass. When the object is cloned
over the user-specified range of $H$, these apparent magnitudes are
adjusted accordingly. When evaluating a specific metric ({\it e.g.}
the number of observations obtained for each orbit in the sample for a
range of $H$ magnitudes), the desired SNR cutoff can be specified and
calculated, including trailing losses and the $5\sigma$ limiting magnitude for each visit.

\begin{figure}
\centering
\includegraphics[width=0.5\textwidth]{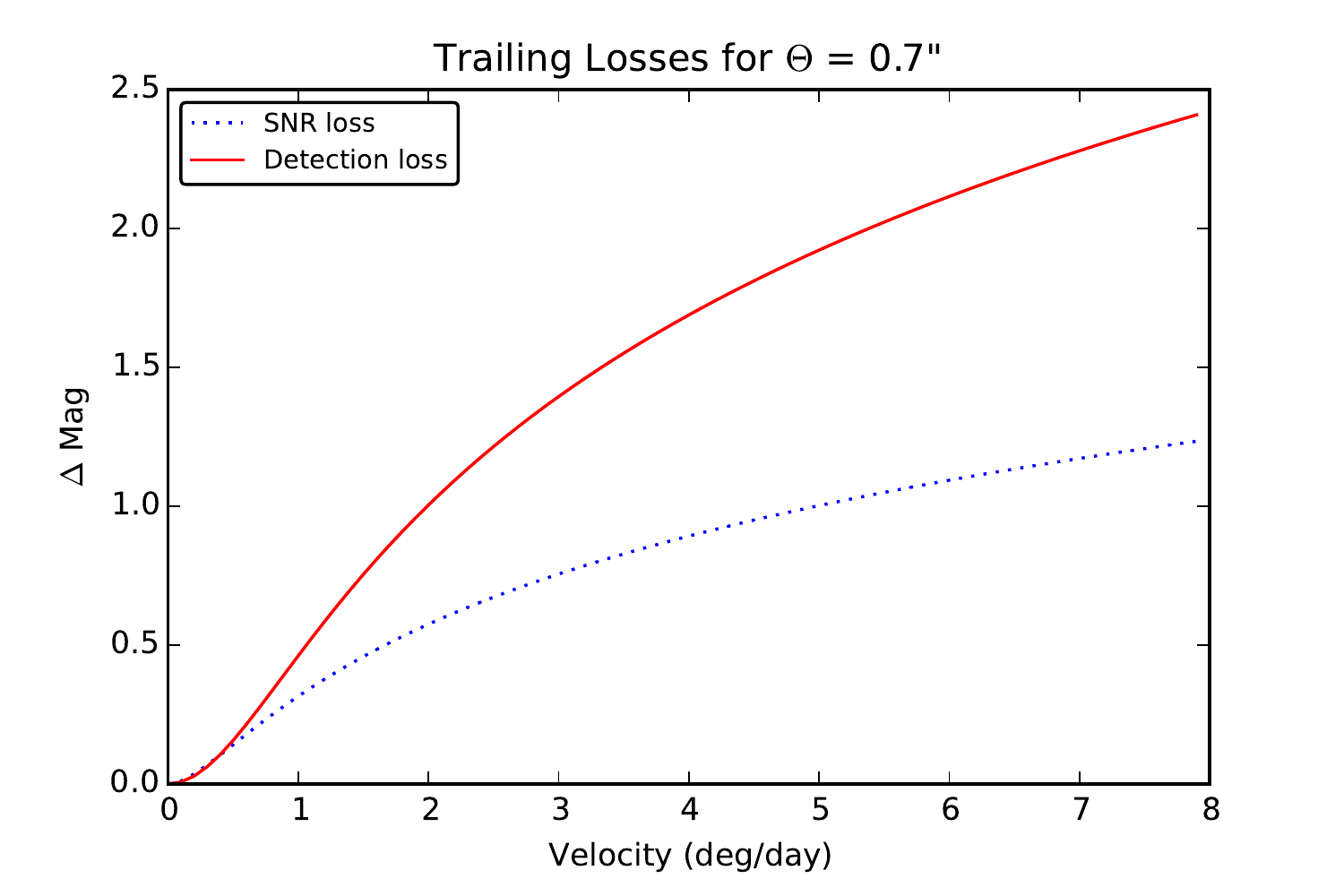}
\caption{Trailing losses for 30 second LSST visits, assuming seeing of
  0.7''. The dotted line shows SNR trailing losses, the solid line
  indicates  detection trailing losses. With software improvements
  detection losses can be mitigated.
\label{trailinglosses}}
\end{figure}

Trailing loss estimates are provided by MAF. Trailing losses occur
whenever the motion of a moving object spreads their light over a
wider area than a simple stellar PSF. There are two aspects
of trailing loss to consider: simple SNR losses and detection losses.
The first is simply the degradation in SNR that occurs (relative to a
stationary PSF) because the trailed object includes a larger number of
background pixels in its footprint. This will affect photometry and
astrometry, but typically doesn't directly affect whether an object is
detected or not. The second effect (detection loss) is not related to
measurement errors but does typically affect whether an object passes
a detection threshhold. Detection losses occur because source
detection software is optimized for detecting point sources;
a stellar PSF-like filter is used when identifying sources that pass
above the defined threshhold, but this filter is non-optimal for
trailed objects. This can be mitigated with improved software ({\it
e.g.} detecting to a lower SNR threshhold and attempting to detect
sources using a variety of trailed PSF filters). Both trailing losses can
be fit as:
\begin{eqnarray}
\Delta \, m & = &-1.25 \, log_{10} \left( 1 + \frac{a \, x^2} { 1 + b\,
    x} \right) \\
x & = & \frac{v \, T_{exp}} {24 \, \theta} 
\end{eqnarray}
where $v$ is the velocity (in degrees/day), $T_{exp}$ is the exposure
time (in seconds), and $\theta$ is the FWHM (in arcseconds). For
SNR trailing losses, we find $a = 0.67$ and $b = 1.16$; for
detection losses, we find $a=0.42$ and $b=0$. An illustration of the
magnitude of these trailing loss effects for 0.7'' seeing is given in
Figure~\ref{trailinglosses}. When considering whether a source would
be detected at a given SNR using typical source detection software,
the detection loss should be used.

MAF can also include the details of the camera focal plane layout, as
illustrated in Figure~\ref{focalplane}; detections which would fall into
chip gaps are then removed.

To demonstrate the potential of LSST as a tool for studying the
Solar System, we calculate a variety of metrics for a set of small
body population samples ranging from Potentially Hazardous Asteroids
(PHAs) to TransNeptunianObjects (TNOs). The 2000 sample orbits used for each
population except the PHAs come from the Pan-STARRS Synthetic Solar
System Model \cite{s3m}; the PHA orbits are taken from the Minor
Planet Center\footnote{\url{http://www.minorplanetcenter.net/}} database,
trimmed to the subset of $\approx1500$ objects larger than 1km in
diameter. In all metrics shown here, we then clone these orbits over a
range of $H$ magnitudes, assumed the (larger) detection trailing
loss, included the camera focal plane footprint, and only used
resulting detections above a SNR=5 cutoff.

\begin{figure}[tb]
\begin{minipage}{.5\textwidth}
\begin{center}
\includegraphics[width=0.9\linewidth]{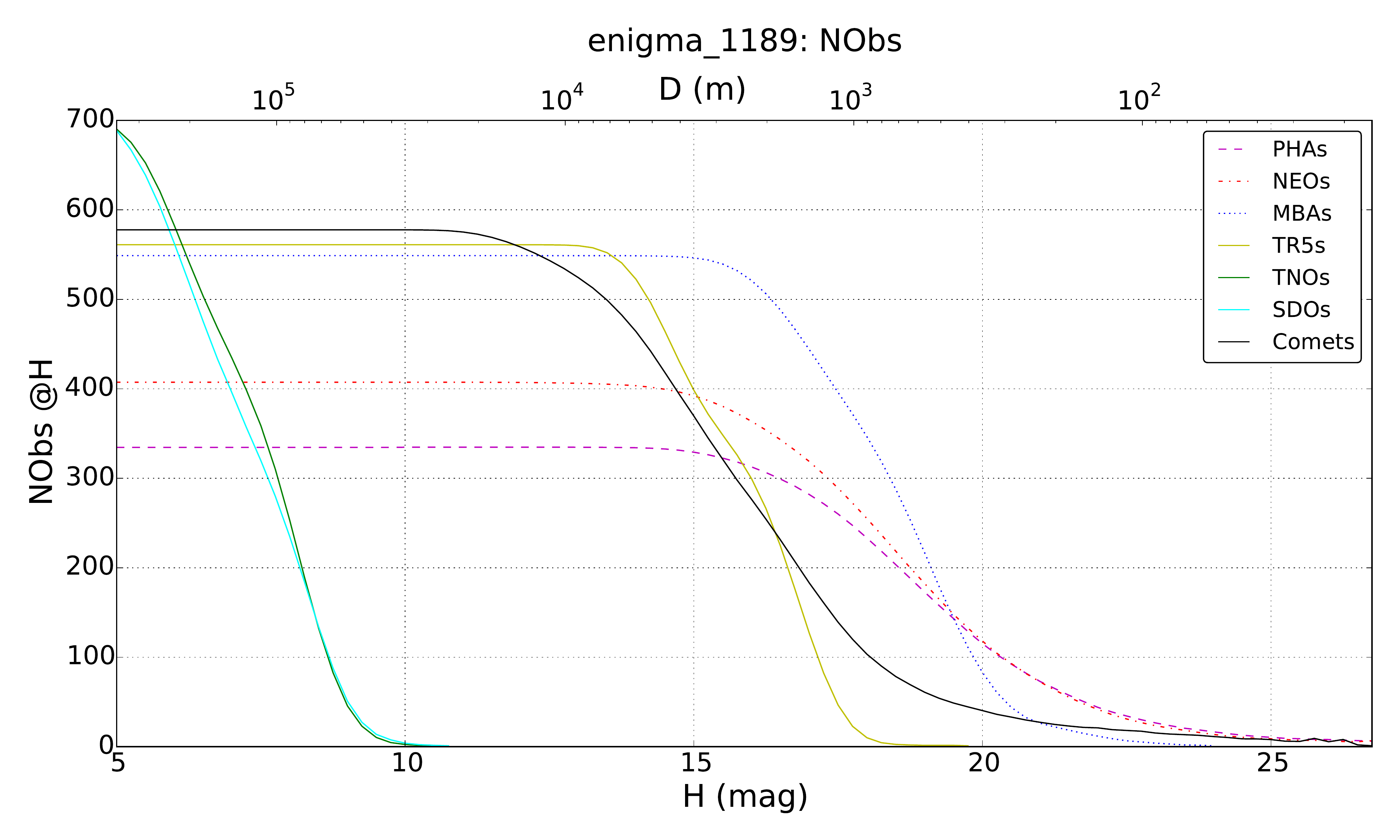}
\captionsetup{width=0.9\linewidth}
\caption{The mean number of observations (per object) for each of our sample small
  body populations, as a function of $H$ magnitude.
\label{nobs}}
\end{center}
\end{minipage}
\begin{minipage}{.5\textwidth}
\begin{center}
\includegraphics[width=0.9\linewidth]{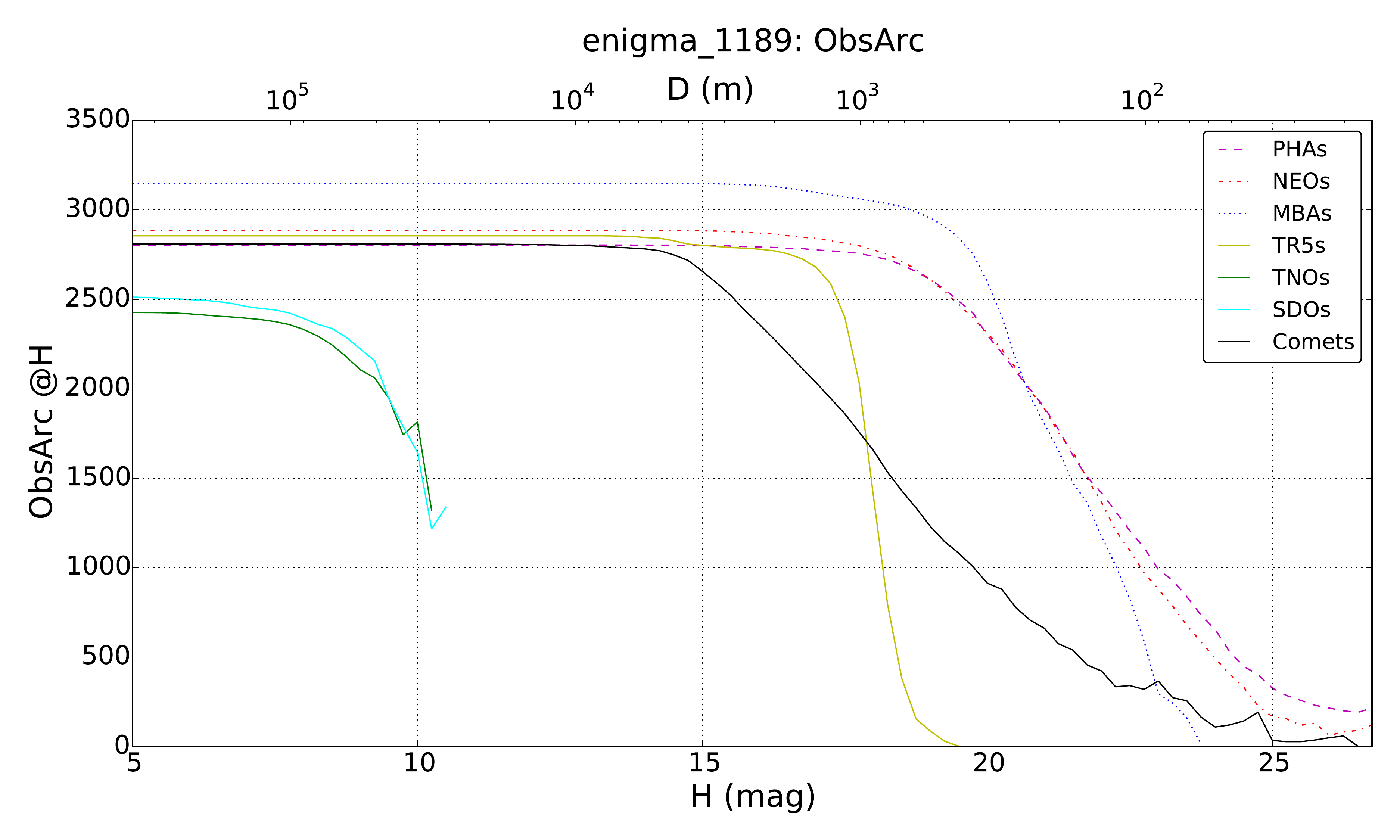}
\captionsetup{width=0.9\linewidth}
\caption{The mean observational arc (in days) for each of our sample
  small body populations, as a function of $H$ magnitude.
\label{arclength}}
\end{center}
\end{minipage}
\end{figure}

First we simply count the total number of observations for each orbit
as a function of $H$; the mean value for all orbits in each population
is shown in Figure~\ref{nobs}. Similarly, we can look at the time of
the first and last observation to get the overall observational arc
length; the mean values  of the observational arc are shown in
Figure~\ref{arclength}.

\begin{figure}[tb]
\begin{center}
\includegraphics[width=0.45\linewidth]{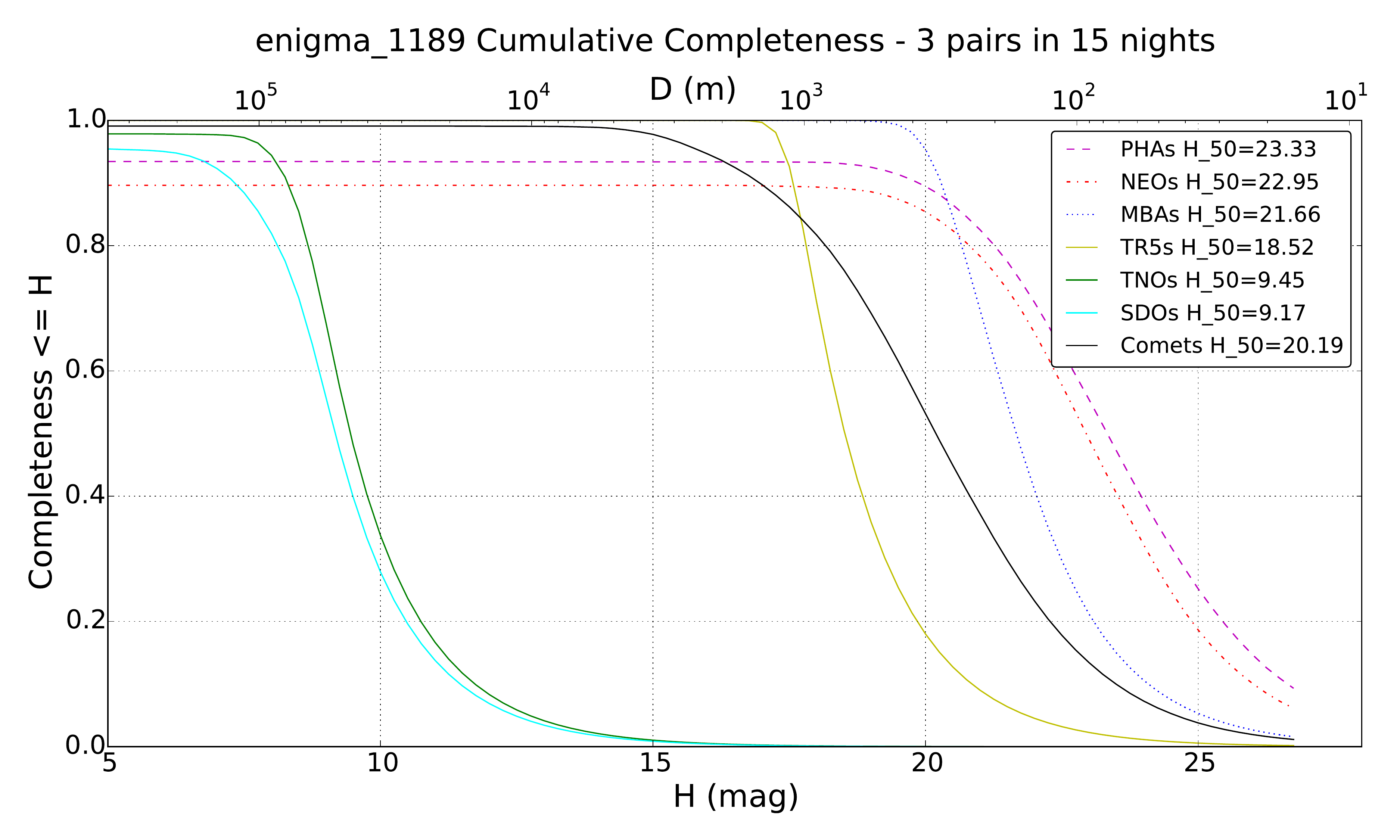}
\includegraphics[width=0.45\linewidth]{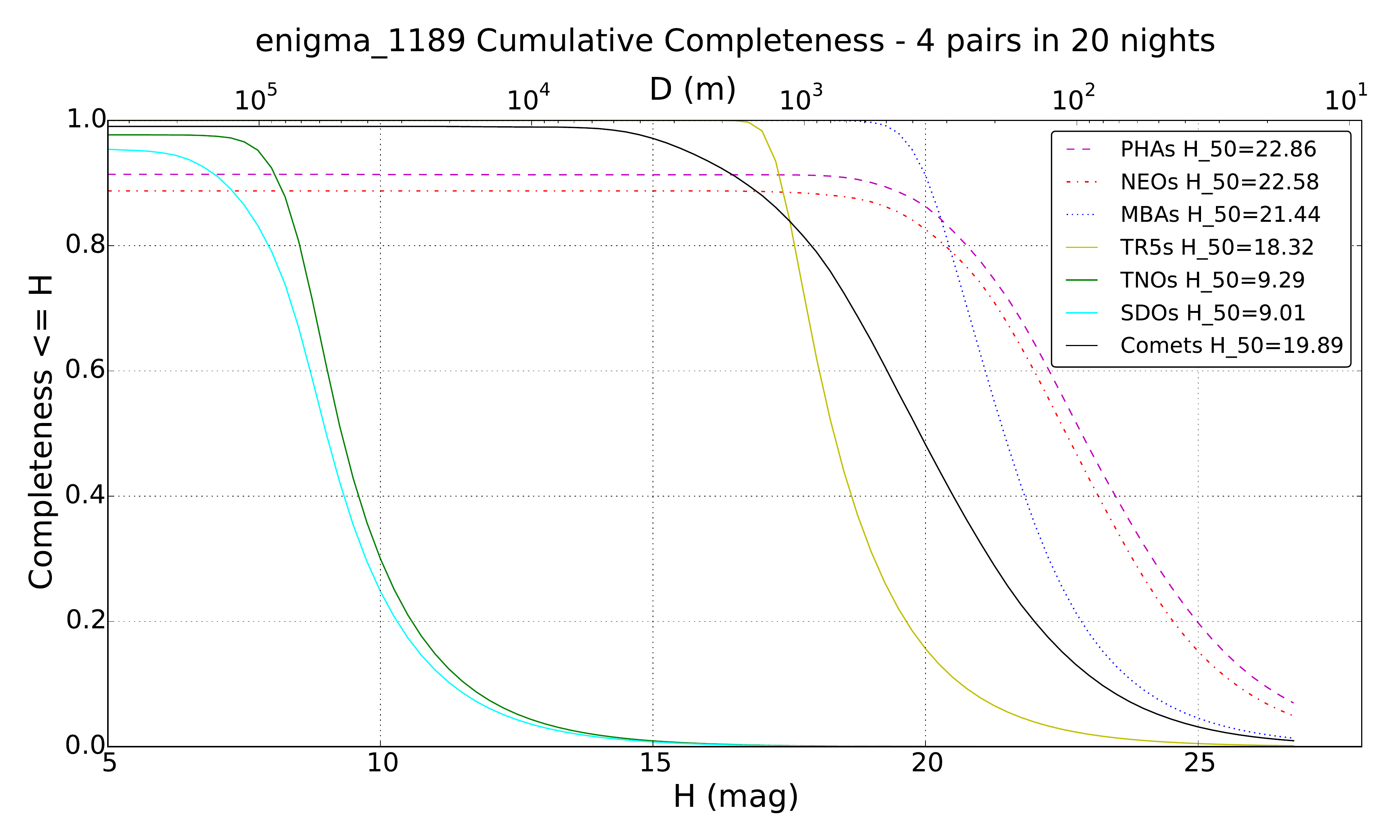} \\
\includegraphics[width=0.45\linewidth]{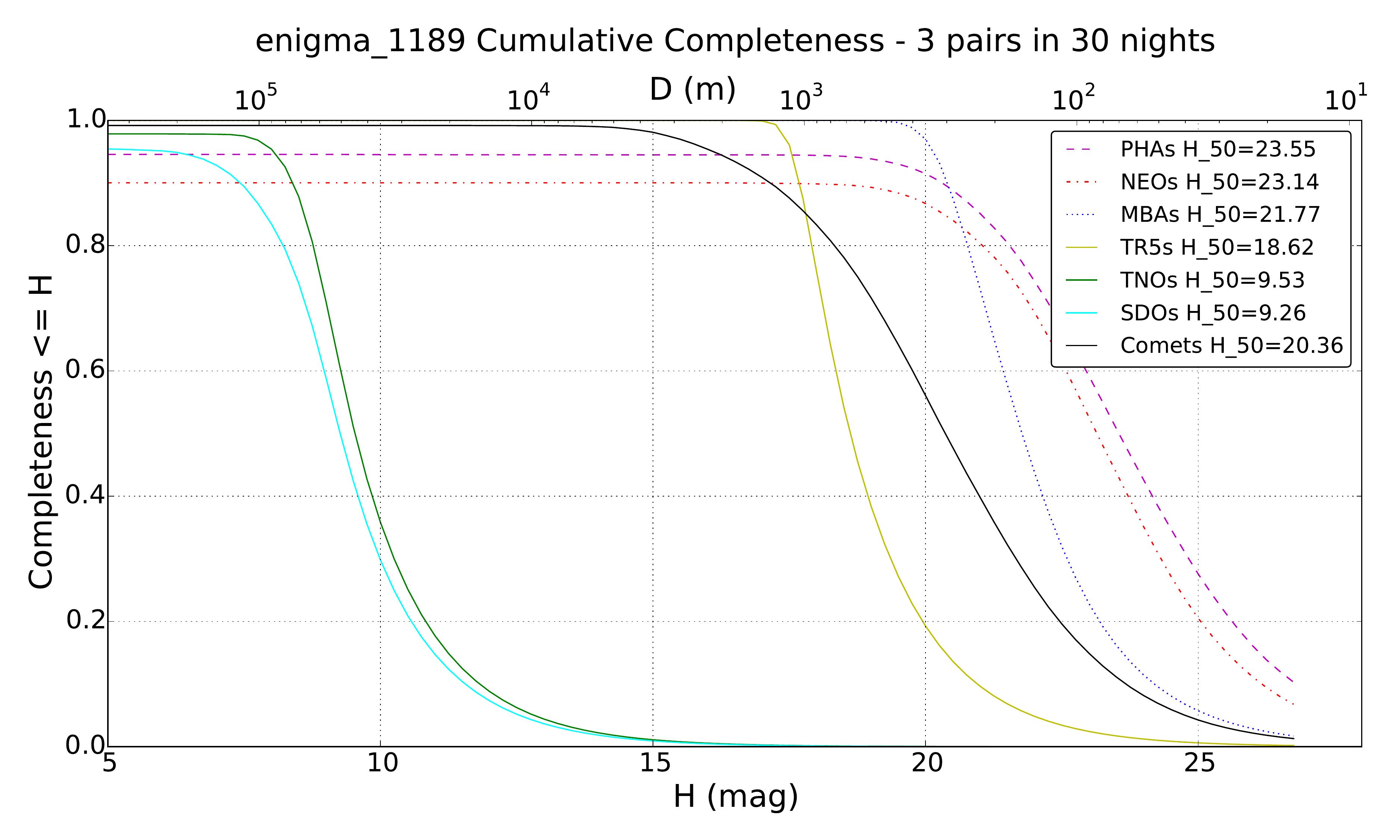}
\includegraphics[width=0.45\linewidth]{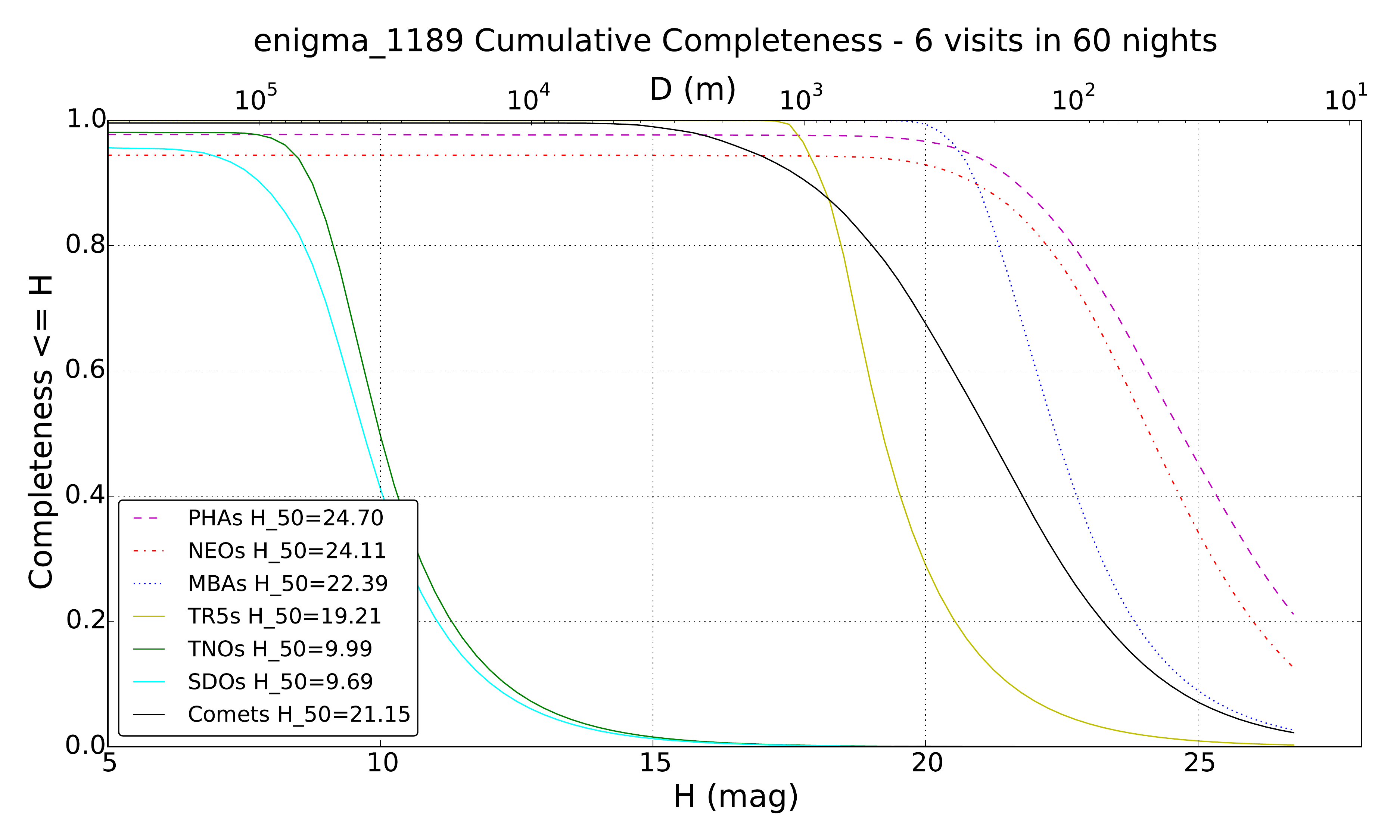}
\caption{Cumulative completeness calculated using various discovery
  criteria as outlined in the text, for each of our sample small body
  populations. $H_{50}$ in the plot legends indicates the $H$ value where
  the cumulative completeness falls to 50\% of the overall population.
\label{completeness}}
\end{center}
\end{figure}

We also calculate the number of `discovery opportunities' available
for each object and use this to calculate the overall completeness
across the population (counting an object as `discovered' if it had at
least one discovery opportunity). The definition of a discovery
opportunity can be varied by the user, but here we look at a variety
of cases. First, the current basic MOPS requirement: detections on
three different nights within a window of 15 nights, with detections
in at least two visits per night separated by less than 90
minutes. Second, an extended MOPS requirement intended to be more
rigorous while still nominally matching the typical observing pattern
in our OpSim survey: detections on four different nights within a
window of 20 nights, with at least two visits per night. Third, a
relaxed discovery criteria intended to demonstrate the effect of
improving MOPS software: 3 nights within 30 days, again with at least
2 visits per night. Finally, a `magic' discovery criteria intended to
get an idea of the upper limit for detection if linking software is
not a constraint: 6 observations in 60 nights. The cumulative
completeness (completeness for $H \le X$) is calculated by multiplying
the differential completeness values by a power law
($\alpha=0.3$). The results are shown in Figure~\ref{completeness},
including the value $H_{50}$, corresponding to the $H$ value where the
cumulative completeness drops to 50\% of the overall population. It
can be see that these varying discovery scenarios have the largest
effects on the PHA and NEO populations. With more rigorous
requirements, the $H_{50}$ values are increased by a few to several
tenths of a magnitude; with relaxed requirements, these values are
pushed faintward by a few tenths. The peak completeness levels change
by a few percent for PHAs and NEOs only. This suggests that even with
the basic cadence LSST is doing fairly well at discovering moving
objects; with improvements in the cadence (probably some version of a
rolling cadence to concentrate more visits into a given chunk of time)
it could do better; and money spent on improving linking software,
even by a relatively modest amount, directly leads to improvements in
completeness.

\begin{figure}[bt]
\begin{minipage}{.5\textwidth}
\begin{center}
\includegraphics[width=0.9\linewidth]{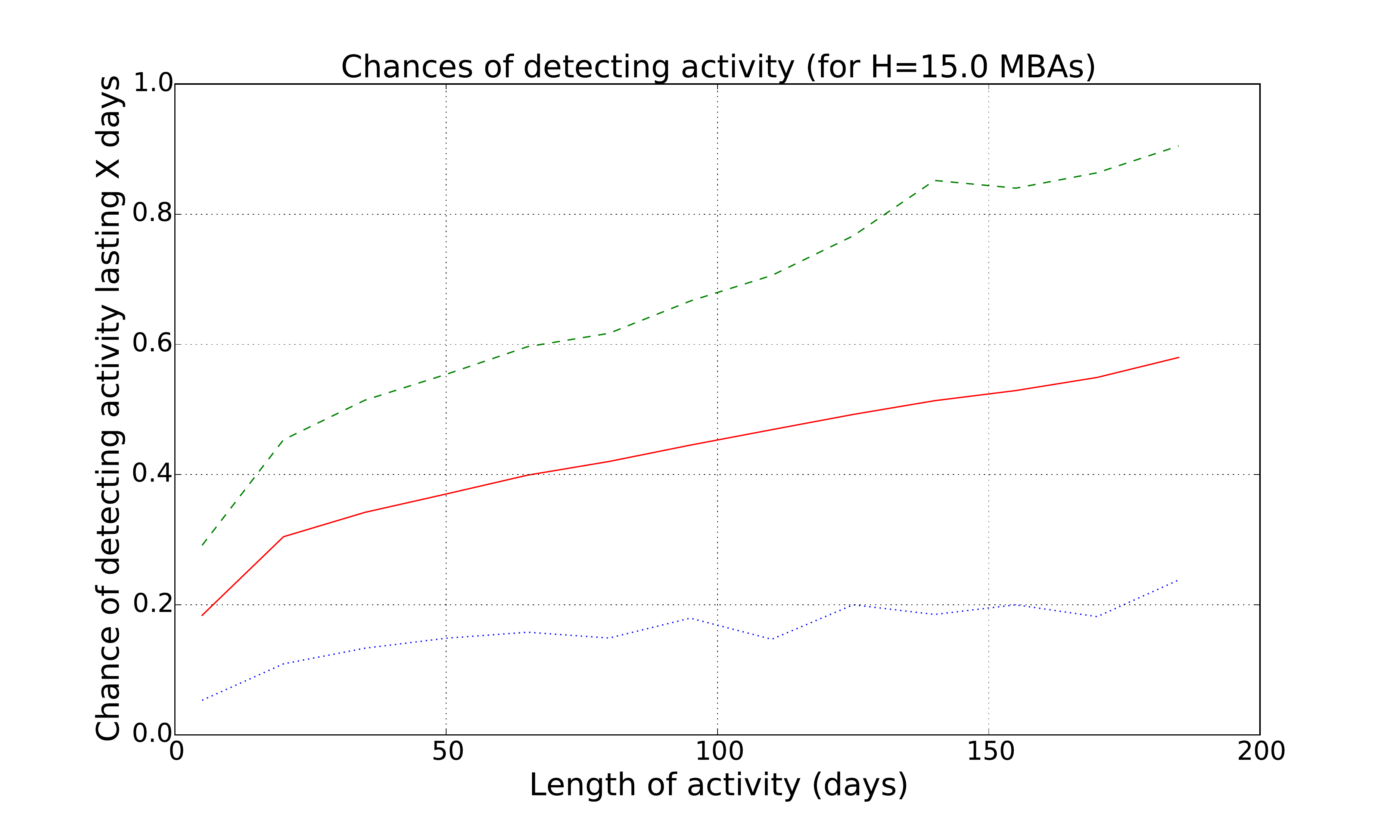}
\captionsetup{width=0.9\linewidth}
\caption{The likelihood of detecting activity which lasts at least a
  given amount of time (in days); mean (solid)
  probabilities across each of the sample populations, as well as maximum
  (dashed) and minimum (dotted) probabilities for individual objects
  within the sample.
\label{activityTime}}
\end{center}
\end{minipage}
\begin{minipage}{.5\textwidth}
\begin{center}
\includegraphics[width=0.9\linewidth]{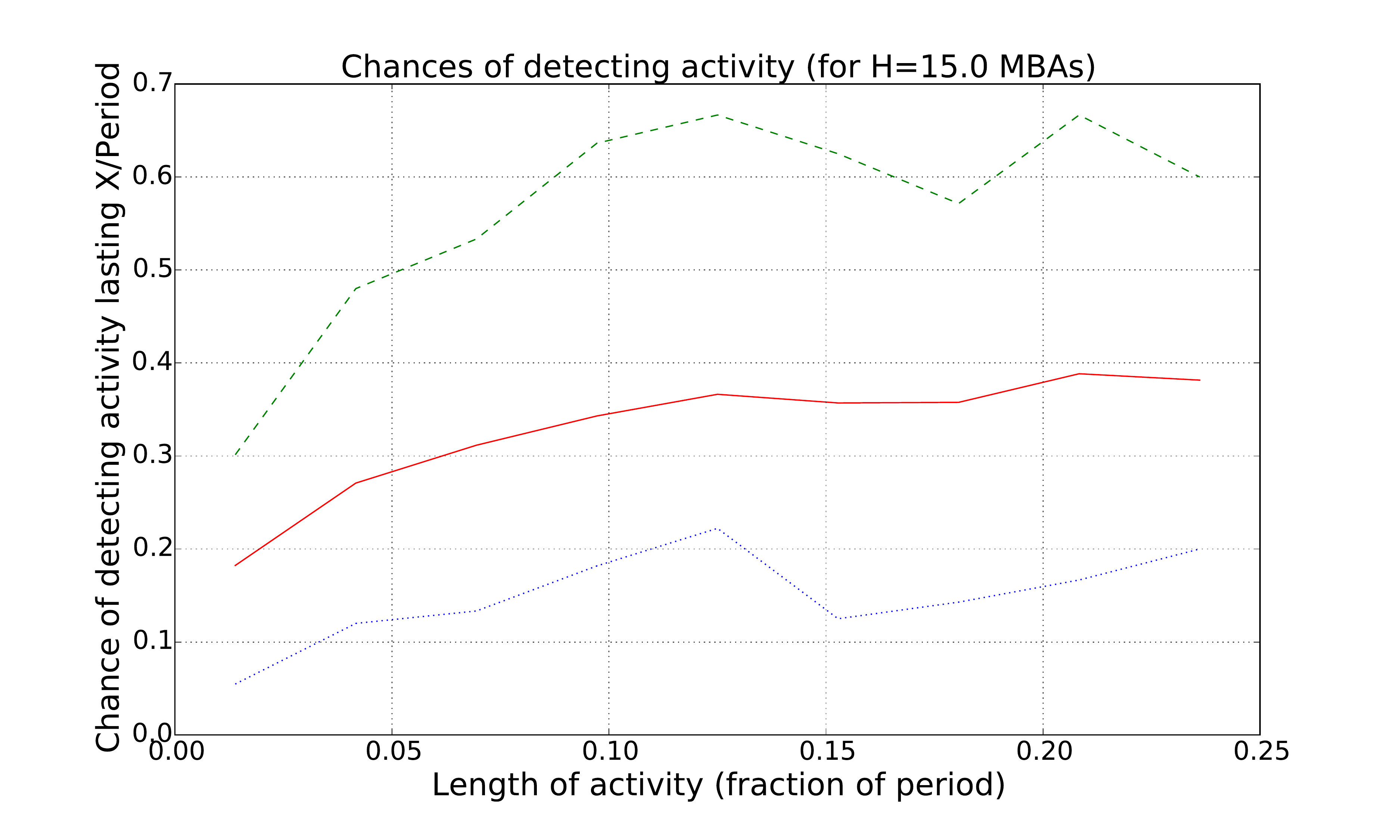}
\captionsetup{width=0.9\linewidth}
\caption{The likelihood of detecting activity which lasts at least a
  given fraction of the orbital period of the object; mean (solid)
  probabilities across each of the sample populations, as well as maximum
  (dashed) and minimum (dotted) probabilities for individual objects
  within the sample.
\label{activityPeriod}}
\end{center}
\end{minipage}
\end{figure}

To demonstrate a more specialized MAF metric, aimed at evaluating the
capability of LSST to help determine the source of activity in active
asteroids, we also present the result of a `activity detection'
metric. Here we take the detections of each object after applying all
the focal plane and SNR cuts and bin the times of these detections
according to either time since the start of the survey (interesting if
activity is due to collisions and thus random) or by time relative to the
period of the object (the interesting timescale if activity is
periodic and associated with the object's orbit). We
then have calculated the probability of observing the object on a
given timescale, and if we assume LSST DM will provide enough
information to identify activity if it is present, this is equivalent
to the probability of detecting activity on that given
timescale. Repeating this exercise over a variety of timescales, we
come up with the probabilities of detecting activity shown in
Figures~\ref{activityTime} and \ref{activityPeriod}.

\section{Discovering moving objects with MOPS}

The discovery requirement with the LSST Moving Object Pipeline System
(MOPS) is: detections on least three separate nights within a 15 night
window, with at least two detections (visits) in each night, separated
by 15 to 90 minutes. The detections within each night are joined into
tracklets, and then tracklet detections from multiple nights are
linked into tracks, which can then be fed to orbit determination
algorithms to filter true from false linkages.

Tests of prototype LSST MOPS with these requirements and an additional
constraint that the velocity on each night must be below 0.5
degrees/day, running on modest hardware (16 cores, $<20$ GB of RAM)
showed that in the absence of noise, moving object detections at the
full depth and cadence of LSST could be easily linked together into
tracks. These tests were repeated with increasing levels of random
noise in the input detection lists. At a ratio of 4:1 noise:real
detections, MOPS was still successful in creating tracks from the
input catalogs; although the software compute requirements (runtime
and memory usage) increased, there was no significant loss in terms of
found
objects\footnote{\url{https://github.com/lsst/mops_daymops/blob/master/doc/report2011/LDM-156.pdf}}.

The noise, or false positive rate of difference image detections, is a
crucial consideration for MOPS. Reducing the false positive rate
places high quality demands on the camera and optical system to reduce
defects and ghosting and on the difference image software to reduce
artifacts. Prototype LSST sensors have been delivered and are testing
within specification and are cosmetically clean. Amp-amp crosstalk is
well within specifications, and CCD-CCD crosstalk is too small to be
measurable. Tests have not shown any measurable charge persistence
under expected operating conditions. The optical system has been
extensively modeled and has extremely small optical ghosting over the
full focal plane. LSST is investing a significant amount of effort
into its difference image software, both for moving object detection
and for the purposes of the Alert pipeline. Existing surveys such as
the Palomar Transient Factory and the Dark Energy Survey are already
using advanced difference image pipelines on cosmetically clean and
well characterized systems to achieve false positive rates of 13:1
(noise:real); with the addition of machine-learning algorithms to
filter artifacts, these pipelines can achieve rates of 1:3 noise:real
detection rates (\cite{goldstein}). Existing surveys are already achieving false-positive
rates within the acceptable range of our prototype MOPS.

Work is ongoing to understand the limitations and capabilities of
MOPS, and the prototype MOPS software will also be improved
prior to survey operations.

\section{Conclusion}

The catalogs of minor planets that will come from LSST over its
lifetime have enormous potential for planetary science. Small body
populations throughout the Solar System will see an increase of 10-100
times more objects than currently known, including Earth minimoon,
irregular satellite, and cometary populations. Many of these new
objects will have large numbers of observations over the course of
several years, in multiple filters, allowing for scientific
characterization of the physical properties of these populations.

LSST provides simulation tools (OpSim and MAF) to assess the impact of
the survey strategy on specific science goals. We encourage feedback
from the community, especially in terms of metrics, to help maximize
the scientific return of LSST. Further development and evaluation of
LSST DM pipelines is ongoing, particularly in the areas of difference
imaging and MOPS. First light for LSST is in 2020, with survey
operations starting 2022.


\end{document}